\documentclass{article}
\usepackage{graphicx}
\usepackage[labelformat=empty]{caption}
\usepackage{amsmath}
\begin{document}
\textbf{
\begin{center}
Reduction of Maximum Flow Network Interdiction Problem from The Clique Problem 
\end{center}
}
\begin{center}
$Pawan Tamta^a,B. P. Pande^b,H.S.Dhami^c$
\end{center}
\begin{center}
a)Department of Mathematics, S.S.J Campus Almora,Kumaun University, Uttarakhand, India,pawantamta0@gmail.com.
\\b)Department of Information Technology, S.S.J Campus Almora,Kumaun University, Uttarakhand, India,bp.pande21@gmail.com.
\\c)Department of Mathematics, S.S.J Campus Almora,Kumaun University, Uttarakhand, India,profdhami@rediffmail.com.
\end{center}
\begin{center}
\textbf{Abstract}
\end{center}
Maximum Flow Network Interdiction Problem (MFNIP) is known to be strongly NP-hard problem. We solve a simple form of MFNIP in polynomial time. We review the reduction of MFNIP from the clique problem. We propose a polynomial time solution to the Clique Problem.
\textbf{\\key words}
\\Network Flows, Reduction, Polynomial time solution, NP-hard
\begin{center}
\textbf{1. Introduction}
\end{center}
The maximum flow network interdiction problem (MFNIP) takes place on a network with a designated source node and a sink node. The objective is to choose a subset of arcs to delete, without exceeding the budget that minimizes the maximum flow that can be routed through the network induced on the remaining arcs. The study of MFNIP in particular originates from the Cold War. Now interdiction problems have many applications, including coordinating tactical air strikes [13], combating drug trafficking [16], controlling infections in a hospital [2], chemically treating raw sewage [14], and controlling floods [15].
From mid nineties to now, efforts have been made to develop some effective algorithms for MFNIP. Initially some naive algorithms were developed for interdiction problem such as a branch-and-bound strategy for general graph [7], and methods of varying quality for inhibition of s-t planar graph(planar graphs with both the source and sink on the outer face) [13].
Later in nineties efforts were made to categorize the problem and some polynomial time algorithms were developed on planar graphs for MFNIP. 
In 1993 Phillips [14] proved MFNIP as weakly NP Complete for planar graphs. At the same time Wood [16] introduced the Integer Linear Program (ILP) for MFNIP and proved it strongly NP Hard problem. 
Ricardo A. Collado et. al mentioned in Rutcor Research Report [6] that even the special case(of MFNIP),where the cost of arc removal is the same for each arc (CMFNIP) is known to be strongly NP-hard. It admits a very simple integer programming formulation [16]. A number of valid inequalities are known for this IP, but the integrality gap is still large [1]. The approximability of this problem is still unknown, with no positive or negative results in the literature. Rutcor Research Report [6] further envisages the recent results in the theory of Stackelberg games [3, 4, 12], which suggests that most of the network interdiction models are in fact APX-hard. Inapproximability bounds with a constant factor are known for shortest path interdiction problems,not for network flow interdiction problems. In this paper we concentrate on Cardinality Maximum Flow Network Interdiction Problem (CMFNIP). CMFNIP is also known as k-most vital arc problem [15].CMFNIP is a special case of MFNIP with the restriction that interdiction cost for every arc is same [16]. Therefore in CMFNIP, we have to interdict the given number of arcs. CMFNIP is also strongly NP-hard problem [16].
Recently Altner et al [1], developed two valid inequalities namely Source to Node path inequality and Node to Sink path inequality for linear programming relaxation of CMFNIP. Altner [1] showed that, even when strengthened by valid inequalities the integrality gap of the standard integer program for CMFNIP is not bounded by a constant. 
In this paper we consider a simpler interdiction problem and name it P-CMFNIP. P-CMFNIP is the problem having reduction from the clique problem as shown by Wood [16]. Wood [16] showed MFNIP strongly NP-complete problem, based on this reduction.
In the literature the reduction given by Wood [16], is the only proof to show MFNIP NP-complete.
Altner et.al [1] presented R-Interdiction Covering Problem (RIC). P-CMFNIP has a simple reduction to RIC. Therefore Altner et.al forwarded the proof of Wood [16] to prove RIC NP-complete problem.
In this paper we review the reduction used by Wood [16] and forwarded by Altner et. al [1]. We solve P-CMFNIP in polynomial time. Clique Problem has a reduction to P-CMFNIP therefore we solve Clique Problem in polynomial time.
In section 2 we review some notations and ILP proposed by Wood [16] and Altner [1]. In section 3 we define P-CMFNIP. In section 4 we review the reduction of P-CMFNIP form the Clique Problem. In section 5 we propose a linear programming solution to P-CMFNIP. In section 6 we propose a polynomial time algorithm to solve P-CMFNIP. In section 7 we propose a polynomial time algorithm to solve Clique Problem. Section 8 is about conclusion.
\begin{center}
\textbf{2-Preliminaries}
\end{center}
A network is defined as $(N,A)$  where $N$ is the set of nodes and $A$ is the set of arcs. It is assumed that all of networks have a unique source $S\in N$ and a unique sink $t\in N$. Arc that originates from node $u$ and terminates at node $v$ are denoted by $(u,v)$. The $s-t$ cut is referred as  either a set of arcs that disconnects $s$ from $t$ upon their removal, or alternatively, as a bipartition of the nodes where $s$ and $t$ are not in the same partition. An undirected graph is denoted as $(V,E)$ where $V$ is the set of vertices and $E$ is the set of edges, an edge between vertices $u$ and $v$ by $\{u,v\}$ and an arc between node $i$ and $j$ as $(i,j)$. The capacity of every arc $(i,j)$ is denoted by $C_e$.The interdiction cost of any arc $e\in A$ is denoted by $r_e$ and total interdiction budget by $R$.
\\Wood [16] proposed the integer linear program for MFNIP and defined the decision variables as:
\begin{center}
\begin{equation*}
\alpha_v=
\begin{cases}
1&\text{if $v\in$$N$ is on sink side of the cut}\\
0&\text{otherwise}
\end{cases}
\end{equation*} 
\begin{equation*}
\beta_e=
\begin{cases}
1 &\text{if $e$$\in$$A$ is in the cut and is interdicted}\\
0&\text{otherwise}
\end{cases}
\end{equation*}
\begin{equation*}
\gamma_e=
\begin{cases}
1 &\text{if $e$$\in$$A$ is on sink side of the cut and is not interdicted}\\
0&\text{otherwise}
\end{cases}
\end{equation*} 
\end{center}
Integer linear program for complete formulation of MFNIP has been given by Wood[16] as under: 
\begin{equation}
\text{Minimize}\sum{C_e\gamma_e}\tag{2.1}
\end{equation}
\\subject to the conditions
\begin{equation}
\alpha_u-\alpha_v+\beta_(u,v)+\gamma_(u,v)\geq0\tag{2.2}
\end{equation}
\begin{equation}
\alpha_t-\alpha_s\geq1\tag{2.3}
\end{equation}
\begin{equation}
\sum_{e\in A} r_e\beta_e\leq R\tag{2.4}
\end{equation}
\begin{equation}
\alpha_v\in  \{0,1\} ,\forall v\in N\tag{2.5}
\end{equation}
\begin{equation}
\beta_e\in \{0,1\}, \forall e\in A\tag{2.6}
\end{equation}
\begin{equation}
\gamma_e\in \{0,1\},\forall e\in A\tag{2.7}
\end{equation}
Altner [1] obtained the following natural linear programming relaxation for and denoted it as (W-LP), by replacing the binary constraints (2.5), (2.6), (2.7) with non negativity constraints 
\begin{equation*}
\alpha_v\geq0,\forall v\in N
\end{equation*}
\begin{equation*}
\beta_e\geq0,\forall e\in A
\end{equation*}
\begin{equation}
\gamma_e\geq0,\forall e\in A\tag{2.8}
\end{equation}
In order to strengthen W-LP for CMFNIP Altner [1] proposed two inequalities named as Node to sink path inequality and Source to node path inequality. 
Node to sink path inequality
\begin{equation}
(|P_{u-t}|-R)\alpha_u+\sum_{e\in A}(P_{u-t})\gamma_e\geq|P_{u-t}|-R,\forall\alpha\in N,P_{u-t}\in p_{u-t}^R\tag{2.9}
\end{equation}
Where $p_{u-t}^R$ denotes the family of all sets of arc-disjoint $u-t$ paths that contain more than $R$ paths.
\\Source to Node Path Inequality 
\begin{equation}
(R-|P_{s-u}|)\alpha_u+\sum_{e\in A}(P_{u-t})\gamma_e\geq|P_{u-t}|-R,\forall\alpha\in N,P_{u-t}\in p_{u-t}^R\tag{2.10}
\end{equation}
Where $p_{s-u}^R$ denotes the family of all sets of arc-disjoint $s-u$ paths that contain strictly greater than $R$ paths.
In this paper we modify the integer program of Wood [35] and Altner [3] for P-CMFNIP to get rid of $\alpha$ variables. But first we define P-CMFNIP in next section.
\textbf{
\begin{center}
3- The problem P-CMFNIP
\end{center}
}
In this section we define a simpler interdiction problem named as P-CMFNIP (figure 3.1)
\begin{figure}[ht]
\centering
\includegraphics[scale=0.5]{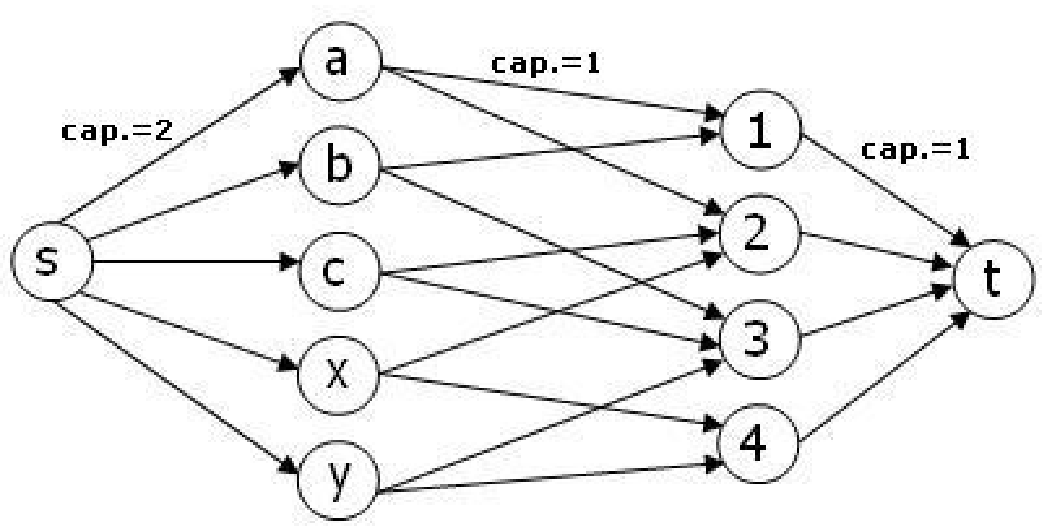}
\caption{figure 3.1}
\label{shape3}
\end{figure}
\\We impose restrictions on MFNIP to make it a simple problem named as P-CMFNIP.
The simplicity of P-CMFNIP lies in the fact that it has only four node sets $V_1$,$V_2$,$V_3$ and $V_4$. Node set $V_1$ has a single node, the source node $s$. Similarly the node set $V_2$ has also a single node $t$ which is the sink node. The nodes of any node set are connected to next node set only. Therefore there are arcs connecting the node set $V_1$ to node set $V_2$ only. Similarly the node set $V_2$ is connected to node set $V_3$ and the node set $V_3$ is connected to node set $V_4$ only. We have no arcs connecting any other combination of the node sets. Further each node in $V_2$ is connected to exactly two nodes in $V_3$. The interdiction cost of every arc is 1. The capacity of every arc connecting node set $V_1$ to node set $V_2$ is 2 units. The capacity of each of the remaining arc is 1 unit. Given the interdiction budget $R$ we are required to interdict arcs connecting the first and second node sets only so that the flow induced in the remaining network is minimum. Furthermore the interdiction budget $R$ can assume the positive integer values only. 
\textbf{
\begin{center}
4 Reduction of the Clique Problem to P-CMFNIP
\end{center}
}
In this section we review the reduction used by Wood [16] in section 3. 
The clique problem (decision) [16] is given as; given an undirected graph $H= (V,E)$ and a positive constant $K$ , does there exists a subgraph of $H$ (complete graph) which is a clique on $K$ vertices? Here $V$ is the set of nodes and $E$ is the set of arcs. Clique of size $K$ is a complete subgraph of $H$ on $K$ vertices i.e. $K\subset V$ such that every two nodes in it are connected by some arc in $E$. \\For a given undirected graph $H=(V,E)$ the reduction given by Wood[16] is as follows:
\begin{figure}[ht]
\centering
\includegraphics[scale=0.5]{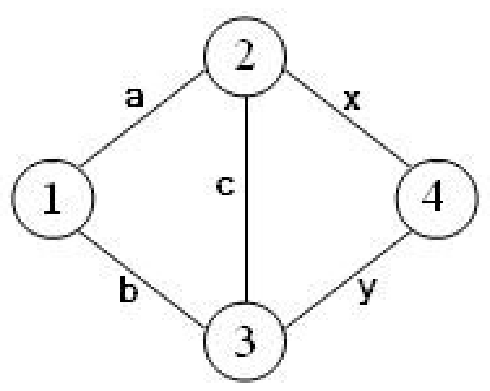}
\caption{figure 4.1}
\label{shape2}
\end{figure}
For each arc in E (figure 4.1) a node is constructed in $V_1$ (figure 3.1). For each node in $V$ (figure 7.1) a node is constructed in $V_2$ (figure 3.1). Every node in $V_1$ is connected to exactly two nodes in $V_2$ (the idea is that one arc connects exactly two nodes). Every node in $V_1$ is connected to a source node and every node in $V_2$ is connected to a sink node. The interdiction cost of every arc is $1$. The arc capacity of every arc connecting source node to $V_1$ has capacity $2$, every arc connecting $V_1$ to $V_2$ has capacity $1$, and every arc connecting $V_2$ to sink node has capacity $1$. Wood [16] proved Lemma 1 and Lemma 2 in section 3 [16] to show that figure 4.1 contains a clique of size $K$ if and only if the interdiction of $R = |E|-C^K_2$ nodes from $V_1$ yields the maximum flow of $K$ units in the remaining network.
Clique problem is NP-complete [11]. P-CMFNIP is the simple version of MFNIP having reduction from the clique problem. Therefore based on this reduction Wood [16] proved that MFNIP is strongly NP-complete problem.
\textbf{
\begin{center}
5-Formulation of the Linear Program for P-CMFNIP
\end{center}
}
In this section we simplify the integer programming solution to MFNIP given by Wood [16].
We relax the integer program and strengthen it by an inequality. We show that a particular optimum solution to the integer program can be decided by this strengthened linear program also.
Wood [16] proved in lemma 2 of section 3 that the maximum flow in the network of figure 3.1 is equal to the number of nodes in node set $V_3$.
Therefore the flow of 1 unit can be interdicted by interdicting any node from the node set $V_3$. Any node from $V_3$ can be interdicted by interdicting all arcs incident to it. These arcs in turn can be interdicted by interdicting respective nodes form $V_2$. Interdiction of any node form $V_2$ is same as the interdiction of respective arc connecting source node to that node.
\begin{figure}[ht]
\centering
\includegraphics[scale=0.5]{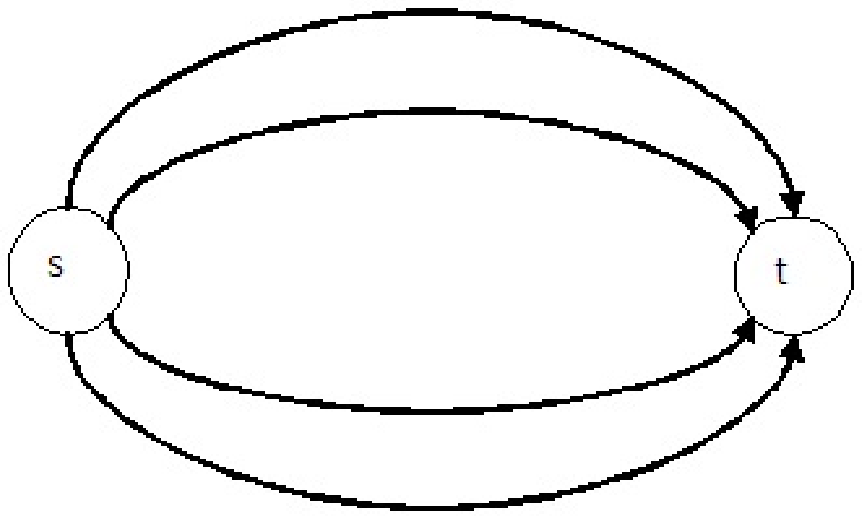}
\caption{figure 5.1}
\label{shape2}
\end{figure}
Therefore P-CMFNIP can be presented as the problem of interdicting nodes from $V_3$. Each node having flow of 1 unit and the interdiction cost equal to the incoming degree of that node (figure 3.1 and figure 5.1). Every arc in figure 5.1 is in the cut so we do not use $\alpha$ variables as defined in section 2.
Further we observe in figure 3.1 that interdiction costs of node $1$ and node $2$ are $2$ and $3$ respectively. But the simultaneous interdiction cost of node $1$ and node $2$ is $4$ instead of $5$. This is because node $a$ is connected to node $1$ and node $2$ both. Each node in $V_2$ represents an edge and each node in $V_3$ represents a node of figure 4.1. Therefore any pair of node from $V_3$ can have only one unique node connected to them from $V_2$. We define the variable $\beta_{i,j}$ for node $i$ and node $j$ from $V_3$, if they are connected to some common node from $V_2$.   
\\Now we modify the integer program based on figure 5.1 for P-CMFNIP as given under:
\begin{equation}
Minimize Z=\sum_{i=1}^{4}\gamma_i\tag{5.1}
\end{equation}
Equation 5.1 is same as equation 2.1 given by Wood [16]
\\Subject to the constraints
\begin{equation}
2\beta_1+3\beta_2+3\beta_3+2\beta_4 - \beta_{1,2}-\beta_{1,3}-\beta_{2,3}-\beta_{2,4}-\beta_{3,4}\leq R\tag{5.2}
\end{equation}
Constraint 5.2 as given above is same as constraint 2.4. Here we have introduced the variables $\beta_{i,j}$. The inclusion of $\beta_{i,j}$ variables facilitates the simultaneous interdiction at marginal cost. For example the interdiction of node $1$ and node $2$ will have interdiction cost $4$ instead of 5. This is possible due to the variable $\beta_{1,2}$.
\\We are not using $\alpha$ variables therefore we remove constraint 2.3. Next we introduce a new set of simple linear constraints for new $\beta$ variables $\beta_{i,j}$
\begin{equation*}
\beta_1+\beta_2-2\beta_{1,2}\geq0
\end{equation*}
\begin{equation*}
\beta_1+\beta_3-2\beta_{1,3}\geq0
\end{equation*}
\begin{equation*}
\beta_2+\beta_3-2\beta_{2,3}\geq0
\end{equation*}
\begin{equation*}
\beta_2+\beta_4-2\beta_{2,4}\geq0
\end{equation*}
\begin{equation}
\beta_3+\beta_4-2\beta_{3,4}\geq0\tag{5.3}
\end{equation}
Constraint set 5.3 is of main concern in this program. It simply says that variable $\beta_{i,j}$ will be active iff both $\beta_i$ and $\beta_j$ are active.
\\Every path of figure 5.1 is in the cut so, constraint 2.2 is expressed as a simple constraint 5.4 as given under
\begin{equation}
\beta_i + \gamma_i \geq 1 \forall i\in A\tag{5.4}
\end{equation}
The integer programming constrains 2.5, 2.6, 2.7 are expressed by constrains 5.5 and 5.6 as given under
\begin{equation}
\gamma_i \in\{0,1\}\tag{5.5}
\end{equation}
\begin{equation}
\beta_i\in\{0,1\}, \beta_{i,j}\in\{0,1\}\tag{5.6}
\end{equation}
Equations 5.1, 5.2, 5.3, 5.4, 5.5 and 5.6 provide the modified integer programming solution to P-CMFNIP. We denote this program by IP.
In order to formulate the linear program for P-CMFNIP we relax the integer program (IP) by replacing the constraints 5.5 and 5.6 by the following constraints
\begin{equation}
\gamma_i \geq0\tag{5.7}
\end{equation}
\begin{equation}
\beta_i\geq0, \beta_{i,j}\geq0\tag{5.8}
\end{equation}
Next we develop a simple linear inequality to strengthen the relax program 
\begin{equation}
\sum_{i}\gamma_i\geq K\tag{5.9}
\end{equation}
We denote the strengthened linear program given by equations 5.1, 5.2, 5.3, 5.4, 5.7, 5.8, 5.9 by SLP. Next we prove a lemma.
\textbf{\\Lemma: the optimum solution to SLP is $K$ for interdiction budget $R=|E|-C_2^K$ if and only if there exist a clique of size $K$ in figure 4.1} 
\\\\Proof: Referred to lemma 1 in section 3 by Wood [16], for interdiction budget $R=|E|-C_2^K$ the optimum solution to IP for P-CMFNIP is $K$ if and only if there exist a clique of size $K$ in figure 4.1. It is obvious that if there does not exist a clique of size $K$ then the optimum solution to IP is always greater than $K$. Flow in the network of figure 3.1 is equal to the number of nodes in $V_3$ which can have integer values only.
Suppose we do not have a clique of size $K$ in figure 4.1. The optimum solution to IP is greater than $K$ in this case. Suppose the optimum solution to IP is $K+1$. SLP is the relaxation of IP therefore it will try to lower down the value. The value cannot be lowered to $K$ as in that case we need some additional $\beta$ variable having value 1 which violates the inequality 5.2.
Hence SLP will have the optimum solution $K$ for interdiction budget $R=|E|-C_2^K$ if and only if the figure 4.1 contains a clique of size $K$. 
\\Now we propose the general linear programming formulation for P-CMFNIP (based on SLP) as given under:
\begin{equation*}
Minimize\sum_{i}\gamma_i
\end{equation*}
Subject to the constraitns
\begin{equation*}
\sum_{i}r_i\beta_i -\beta_{j,k}\leq R
\end{equation*}
Where the variables $\beta_{j,k}$ are as defined in IP.
\begin{equation*}
\beta_i+\beta_j-2\beta_{i,j}\geq0
\end{equation*}
\begin{equation*}
\gamma_i+\beta_i\geq1\forall i
\end{equation*}
\begin{equation*}
\gamma_i \geq0
\end{equation*}
\begin{equation*}
\beta_i\geq0, \beta_{i,j}\geq0
\end{equation*}
\begin{equation*}
\sum_{i}\gamma_i\geq K
\end{equation*}
\textbf{
\begin{center}
6-Polynomial time algorithm for P-CMFNIP (Decision)
\end{center}
}
In this section we propose a polynomial time algorithm "Poly-MFNIP" for P-CMFNIP (Decision) based on the linear program (SLP) proposed in section 5.
In the decision version of P-CMFNIP we have to decide whether it's possible to interdict arcs  within the interdiction budget $R$ so that the maximum flow in the network after interdiction is $K$.
\\The algorithm is given as under
\textbf{\\Step 1-} For given interdiction budget $R$ and given constant $K$, SLP (proposed in section 5) is solved
\textbf{\\Step 2-} If the solution to SLP is $K$ then the given network can have a maximum flow $K$ for interdiction budget $R$, else it cannot have the maximum flow $K$ for interdiction budget $R$. 
Step 1 of the algorithm involves the solution of the linear program which is solvable in polynomial time [9,10]. Step 2 being a simple if/else statement, is decided in polynomial time. Therefore the Poly-MFNIP runs in polynomial time.
\textbf{
\begin{center}
7- Polynomial time solution to the clique problem(Decision)
\end{center}
}
In this section we propose a polynomial time algorithm to solve clique problem (decision). 
The problem is stated as; given an undirected graph H= (V,E), we have to find whether there exist a clique of size $K$.
\\The algorithm is based on linear programming formulation (SLP) mentioned in section 5. For that purpose we assign $\gamma$ and $\beta$ variables to each node of given undirected graph $H$.
Then the linear programming formulation for clique problem is given by SLP. In constraint 5.2, the coefficient of the variable $\beta_i$ is the degree of that node. The variable $\beta_{i,j}$ is used for nodes $i$ and $j$ if they are connected by some arc.  
The algorithm is expressed as under
\textbf{\\Step1-} Given an undirected graph SLP is solved for given constant $K$ and $R= |E| - C_2^K$
\textbf{\\Step2-} If the optimum solution to SLP is $K$ then the graph has a clique on $K$ vertices, else it cannot have a clique on $K$ vertices.
Both steps run in polynomial time as shown in section 6.
\textbf{
\begin{center}
7.1-Polynomial time solution to The Maximum Clique Problem (Optimization)
\end{center}
}
The optimization version of the Clique Problem is known as the Maximum Clique Problem. The problem is stated as; given an undirected graph H= (V,E), we have to find the complete subgraph of H of maximum size. Simply speaking we have to find a clique of maximum size.
The maximum number of vertices in a clique are $|E|$, therefore the algorithm runs as follows
\textbf{\\Step1-} SLP is computed for given value $K=|E|$ and interdiction budget $R= |E|- C_2^K$.  
\textbf{\\Step2-} If the optimum solution to SLP is $K=|E|$ then the graph has a clique on $K=|E|$ vertices, else take $K=|E|-1$ and go to step 1.
\\A clique of size less than 2 in any undirected graph is not possible. Therefore in the loop of step 1 and step 2 the number of efforts cannot exceed $|E|-2$ which is polynomial in $|E|$. Step 1 and Step 2 run in polynomial time as shown in section 6.
\textbf{
\begin{center}
8- Conclusion
\end{center}
}
We have shown that P-CMFNIP can be solved in polynomial time. Therefore harness of MFNIP cannot be decided merely on the basis of reduction mentioned in section 4. Further based on the reduction in section 4 we can have a polynomial time solution to the clique problem also. 
\textbf{
\begin{center}
9. References.
\end{center}
}
$[1]$ Douglas.S.Altner, Ozlem Eegun and Nelson A Uhan, The Maximum Flow
Network Interdiction Problem, valid inequalities, integrality gaps and approximability,
Operations Research Letter 38 (2010),pp 33-38.
\\$[2]$ N. Assimakopoulos, A network interdiction model for hospital infection control,
Computers in Biology and Medicine 17 (1987), pp. 413-422.
\\$[3]$ L. Bingol, A Lagrangian heuristic for solving a network interdiction problem,
Master's thesis, Naval Postgraduate School, 2001.
\\$[4]$ Boyd, S. and Carr, R. (1999). A new bound for the ratio between the
2-matching problem and its linear programming relaxation. Mathematical Programming,
Ser A, 86:499-514.
\\$[5]$ Cook S.A. [1971], "The complexity of theorem-providing procedures", Proc.
3rdAnn. ACM symp. On Theory of Computing, Association for Computing
Machinery, New York, 151-158
\\$[6]$ Ricardo.A. Collado , David Papp, Network Interdiction-Models, Applications,
Unexplored Directions, Rutcor Research Report, RRR 4-2012, January
2012.
\\$[7]$ Ghare, P. M., Montgomery, D. C., and Turner, W. C. (1971). Optimal interdiction
policy for a flow network. Naval Research Logistics Quarterly, 18:37-45.
\\$[8]$ Helmbold, R. L. (1971). A counter capacity network interdiction model.
Technical Report R-611-PR, Rand Corporation, Santa Monica, CA.
\\$[9]$ Khachiyan, L.G., "Polynomial algorithm in linear programming,", Soviet
Mathematics Doklady 20, (1979) pp. 191-194.
\\$[10]$ Karmarkar, N., "A new polynomial-time algorithm for linear programming,"
Combinatorica 4, (1984) pp. 373-395
\\$[11]$ Karp R.M. [1972], "Reducibility among combinatorial problems", in R.E
Miller and J.W. Thatcher (eds.), Complexity of Computer Computations, Plenum
Press, New York, 85-103.
\\$[12]$ Ford, L. R. and Fulkerson, D. R. (1962). Flows in Networks. Princeton
University Press, Princeton, NJ.
\\$[13]$ A.W. McMasters and T.M. Mustin, Optimal interdiction of a supply network,
Naval Research Logistics Quarterly 17 (1970), pp. 261-268.
\\$[14]$ C.A. Phillips, The network inhibition problem, in: Proceedings of the 25th
Annual ACM Symposium on the Theory of Computing, 1993 pp. 776-785.
\\$[15]$ H.D. Ratliff, G.T. Sicilia and S.H. Lubore, Finding the n most vital links
in flow networks, Management Science 21 (1975), pp. 531-539.
\\$[16]$ R.K. Wood, Deterministic network interdiction, Mathematical and Computer
Modelling 17 (1993), pp. 1-18.
\end{document}